\RequirePackage{fixltx2e}

\documentclass[aps,prc,reprint,superscriptaddress]{revtex4-1}

\usepackage{graphicx}
\graphicspath{{./graphics/}}  
\usepackage{amsmath}
\usepackage{array}
\newcolumntype{+}{>{\global\let\currentrowstyle\relax}}
\newcolumntype{^}{>{\currentrowstyle}}
\usepackage{lipsum}
\usepackage{booktabs}
\usepackage{dcolumn}
\usepackage{cellspace}

\usepackage{scrextend}
\usepackage{hyperref}
\newcommand{\comment}[1]{}

\newcommand{\degree}{\ensuremath{^\circ}}
\usepackage{rotating}          

\newcommand{\gtaeq}{\lower 2pt \hbox{$\, \buildrel {\scriptstyle >}\over {\scriptstyle \sim}\,$}}
\newcommand{\ltaeq}{\lower 2pt \hbox{$\, \buildrel {\scriptstyle <}\over {\scriptstyle \sim}\,$}}

\usepackage[export]{adjustbox} 

\begin{document}



\title{Determining the average prompt-fission-neutron multiplicity for
$^{239}$Pu($n$,$f$) \protect\\ via a $^{240}$Pu($\alpha$,$\alpha^{\prime}f$) surrogate reaction}


\author{B. S. Wang}
\email[]{alan2@llnl.gov}
\affiliation{Lawrence Livermore National Laboratory, Livermore, California 94550, USA}

\author{J. T. Burke}
\affiliation{Lawrence Livermore National Laboratory, Livermore, California 94550, USA}

\author{O. A. Akindele}
\affiliation{Lawrence Livermore National Laboratory, Livermore, California 94550, USA}
\affiliation{Department of Nuclear Engineering, University of California, Berkeley, California 94720, USA}

\author{R. J. Casperson}
\affiliation{Lawrence Livermore National Laboratory, Livermore, California 94550, USA}

\author{R. O. Hughes}
\affiliation{Lawrence Livermore National Laboratory, Livermore, California 94550, USA}

\author{J. D. Koglin}
\affiliation{Lawrence Livermore National Laboratory, Livermore, California 94550, USA}

\author{K. Kolos}
\affiliation{Lawrence Livermore National Laboratory, Livermore, California 94550, USA}

\author{E. B. Norman}
\affiliation{Department of Nuclear Engineering, University of California, Berkeley, California 94720, USA}
\affiliation{Lawrence Livermore National Laboratory, Livermore, California 94550, USA}

\author{S. Ota}
\affiliation{Cyclotron Institute, Texas A\&M University, College Station, Texas 77840, USA}

\author{A. Saastamoinen}
\affiliation{Cyclotron Institute, Texas A\&M University, College Station, Texas 77840, USA}


\date{\today}

\begin{abstract}
The average prompt-fission-neutron multiplicity $\bar{\nu}$ is of significance in the areas of
nuclear theory, nuclear nonproliferation, and nuclear energy.  In this work,
the surrogate-reaction method has been used for the first time to indirectly determine
$\bar{\nu}$ for $^{239}$Pu($n$,$f$) via $^{240}$Pu($\alpha$,$\alpha^{\prime}f$) reactions.
A $^{240}$Pu target was bombarded with a beam of 53.9-MeV $\alpha$ particles.
Scattered $\alpha$ particles, fission products, and neutrons were measured with the NeutronSTARS
detector array. Values of $\bar{\nu}$ were obtained for a continuous range of equivalent incident
neutron energies between 0.25--26.25~MeV, and the results agree well with direct neutron measurements.
  
\end{abstract}


\maketitle{}


\section{Introduction}
\label{sec:Introduction}

The average prompt-fission-neutron multiplicity $\bar{\nu}$ following ($n$,$f$) reactions is important to
both basic and applied physics.
In nuclear theory, measurements of $\bar{\nu}$ can be used to validate fission models and
provide constraints on the fission process itself~\cite{Maslov2004}.
In the area of international safeguards and verification, nuclear materials are assayed with passive neutron-multiplicity counting, and here,
$\bar{\nu}$ is needed to determine the amount of neutron-induced fission (or self-multiplication) in the sample \cite{Ensslin1998,Shin2017}.
For proposed nuclear reactor concepts, such as accelerator-driven systems (ADS) and those based on the thorium-uranium cycle,
there is interest in the $\bar{\nu}$ values for short-lived actinides, as the dependence of $\bar{\nu}$ on the incident neutron
energy is important for determining the criticality, safety, and lifetime of these reactors~\cite{Ethvignot2005,Kerdraon2003,Nifenecker2001}. 
In addition, $\bar{\nu}$ for short-lived actinides is also relevant to transmutation of radioactive waste with ADS~\cite{Ethvignot2005,Kerdraon2003,Nifenecker2001}.

Directly measuring $\bar{\nu}$ presents a number of experimental challenges, including producing high-flux
neutron beams and addressing beam-related backgrounds. 
For short-lived actinides, $\bar{\nu}$ data are particularly sparse due to the fact that target fabrication
and high target activity are also issues. 
These challenges can be bypassed with the surrogate-reaction method~\cite{Escher2012},
an indirect measurement technique that has typically been used to determine the cross sections of reactions
that proceed through a highly excited, statistically equilibrated compound nuclear state.
In a surrogate experiment, the desired compound nucleus (CN) is produced using an alternative (``surrogate'')
reaction with a more experimentally accessible or preferable combination of projectile and target nucleus.
The surrogate method has been demonstrated to work well for determining ($n$,$f$) reaction
cross sections of various actinides~\cite{Petit2004,Plettner2005,Burke2006,Escher2006,Lesher2009,Ressler2011};
the values obtained are within $\sim$5--20\% of direct neutron measurements.
The present work extends the applicability of this technique to determining $\bar{\nu}$. 
Benchmarking has been performed by using the surrogate reactions
$^{240}$Pu($\alpha$,$\alpha^{\prime}f$) and $^{242}$Pu($\alpha$,$\alpha^{\prime}f$)  
to obtain $\bar{\nu}$ as a function of incident neutron energy for the reactions $^{239}$Pu($n$,$f$) and
$^{241}$Pu($n$,$f$), respectively, for which direct-measurement data are available.
The results for $^{239}$Pu($n$,$f$) are discussed in this paper, while those for $^{241}$Pu($n$,$f$) can be
found in Ref.~\cite{Akindele2019}.

\section{Surrogate-reaction technique}
\label{sec:SurrogateReactionTechnique}
In the present work, the compound nucleus $^{240}$Pu in the desired reaction

\begin{equation}
\label{eq:DesiredReaction}
	n + ^{239}\text{Pu} \rightarrow ^{240}\text{Pu}^* \rightarrow \text{LF} + \text{HF} + \nu n
\end{equation}
is produced via the surrogate reaction

\begin{equation}
\label{eq:SurrogateReaction}
	\alpha + ^{240}\text{Pu} \rightarrow \alpha^\prime + ^{240}\text{Pu}^* \rightarrow \alpha^\prime + \text{LF} + \text{HF} + \nu n,
\end{equation}
where LF and HF are the light and heavy fission fragments, respectively, and $\nu$ is the prompt-fission-neutron
multiplicity.
Assuming a statistically equilibrated CN,
where the decay is independent of the method of formation~\cite{Bohr1936},
the ($n$,$f$) cross section for an incident neutron energy $E_n$ is given by the following
Hauser-Feshbach~\cite{HauserFeshbach1952,Frobrich1996,ThompsonNunes2009} formula:

\begin{equation}
    \sigma_{n,f}(E_n) = \sum_{J,\pi}\sigma_{n}^{CN}(E_\text{ex},J,\pi)G_{f}^{CN}(E_\text{ex},J,\pi),
\label{eq:HauserFeshbach}
\end{equation}
where $\sigma_{n}^{CN}(E_\text{ex},J,\pi)$ is the cross section for forming
a CN with excitation energy $E_\text{ex}$, angular momentum $J$, and parity $\pi$, and
$G_{f}^{CN}(E_\text{ex},J,\pi)$ is the probability that the CN will fission. 
In the Weisskopf-Ewing limit of Hauser-Feshbach theory, where the decay of the CN
is independent of $J$ and $\pi$, Eq.~\ref{eq:HauserFeshbach} reduces to

\begin{equation}
    \sigma_{n,f}(E_n) = \sigma_{n}^{CN}(E_\text{ex})G_{f}^{CN}(E_\text{ex}).
\label{eq:WeisskopfEwingLimit}
\end{equation}
Analogously, the ($\alpha$,$\alpha^\prime f$) cross section for an incident $\alpha$-particle energy $E_{\alpha}$ is given by

\begin{equation}
    \sigma_{\alpha,\alpha^\prime f}(E_\alpha) = \sigma_{\alpha,\alpha^\prime}^{CN}(E_\text{ex})G_{f}^{CN}(E_\text{ex}).
\label{eq:SurrogateWeisskopfEwingLimit}
\end{equation}
In Eq.~\ref{eq:WeisskopfEwingLimit} and \ref{eq:SurrogateWeisskopfEwingLimit},
$\sigma_{n}^{CN}(E_\text{ex})$ and $\sigma_{\alpha,\alpha^\prime}^{CN}(E_\text{ex})$ are 
the $J\pi$-independent CN-formation cross sections and $G_{f}^{CN}(E_\text{ex})$
is the $J\pi$-independent fission probability of the CN. 
If the Weisskopf-Ewing approximation applies, then ($n$,$f$) and ($\alpha$,$\alpha^\prime f$)
reactions that generate the same CN with excitation energy $E_\text{ex}$ will have
identical values of $G_{f}^{CN}(E_\text{ex})$ and yield the same $\bar{\nu}$. 
The validity of this assumption is tested by comparing the $\bar{\nu}$ values obtained with the
surrogate reaction $^{240}$Pu($\alpha$,$\alpha^{\prime}f$) to those determined from
direct $^{239}$Pu($n$,$f$) measurements.

\section{Experiment}
\label{sec:ExperimentalMethod}
The experiment was performed in Cave 4 of the Texas A\&M University Cyclotron Institute~\cite{Tabacaru2019}.
A $^{240}$Pu target was loaded onto a target wheel~\cite{Lesher2010} located at the center of the NeutronSTARS array~\cite{Akindele2017}
and bombarded with a 100-pA beam of 53.9-MeV alpha particles from the K150 Cyclotron; 4.75 days' worth of data was collected.

\subsection{Targets}
\label{subsec:Targets}
The $^{240}$Pu target was 99.995\%-pure; it was fabricated by first epoxying a 100-$\mu$g/cm$^{2}$-thick natural-carbon foil
to an aluminum frame, and then electroplating plutonium onto the foil surface, covering a circular area 1.90~cm in diameter.
Properties of the target are given in Table~\ref{tab:Pu240TargetProperties}.

The following calibration targets were included in the experiment:
a $^{208}$Pb foil to determine the beam energy; a natural-carbon foil, Mylar ((C$_{10}$H$_{8}$O$_{4}$)$_\text{n}$)
foil, and empty aluminum frame to assess backgrounds due to $\alpha$ interactions with carbon, oxygen and aluminum
in the $^{240}$Pu target. Two phosphor targets were also used for beam alignment and observing the beam-spot size.

\begin{table}[!tb]
\caption{\label{tab:Pu240TargetProperties} Properties of the $^{240}$Pu target
	used in the experiment.  In addition to $^{240}$Pu, a small amount of
	$^{238}$Pu was also present.
	} 
\begin{ruledtabular}
\begin{tabular}{ l d d }
  Property & \multicolumn{1}{c}{\textsuperscript{240}\text{Pu}} & \multicolumn{1}{c}{\textsuperscript{238}\text{Pu}}\\ 
  \hline\noalign{\smallskip}
  Activity ($\mu$Ci) & 67.374(334) & 0.250(3)\\ 
  Weight Percent (\%) & 99.995(718) & 0.00493(6)\\ 
  Thickness ($\mu$g/cm$^2$) & 104.078(528) & 0.00513(6)\\ 
\end{tabular} 
\end{ruledtabular}
\end{table}

\subsection{Apparatus}
\label{subsec:Apparatus}
\begin{figure}[!tb]
   \includegraphics[width=0.5\textwidth]{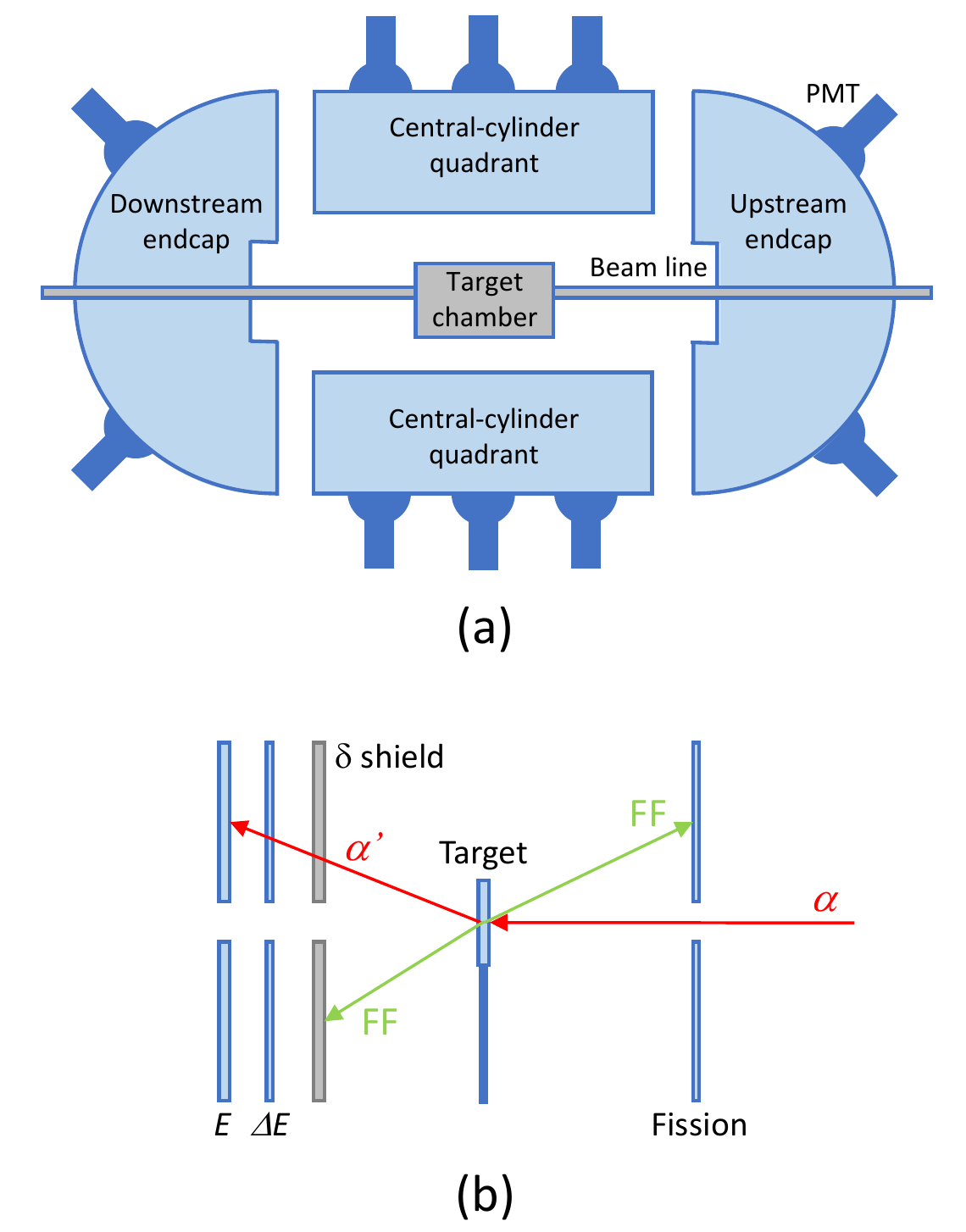}
   \caption
      {
      \label{fig:NeutronSTARS}
      (Color online) Cross-sectional views of (a) the NeutronSTARS detector array and
      (b) the inside of the target chamber (not to-scale); the $\alpha$-particle beam
      travels from right to left.      
      NeutronSTARS consists of a target chamber that sits at the center of a neutron detector.
      The latter is a large tank of gadolinium-doped liquid scintillator segmented into
      six regions: four identical quadrants that make up the central cylinder and two endcaps.
      Three PMTs are attached to each quadrant and four are attached to each endcap.
      The target chamber contains a target wheel, a $\Delta E$-$E$ telescope to measure scattered
      $\alpha$ particles, a fission detector to measure fission fragments (FF), and a $\delta$ shield
      to prevent fission fragments and $\delta$ electrons from hitting the $\Delta E$ detector.
      }
\end{figure}

The NeutronSTARS array is shown in Fig.~\ref{fig:NeutronSTARS}.  
Charged particles, including inelastically scattered $\alpha$ particles from $^{240}$Pu($\alpha$,$\alpha^{\prime}f$)
reactions, were detected with a silicon telescope located 19 mm downstream from the target and
consisting of two Micron S2-type annular silicon detectors
(a 152-$\mu$m-thick $\Delta E$ detector and a 994-$\mu$m-thick $E$ detector)
that were separated by 4~mm.
The energy loss in the two detectors was used for particle identification.
A 4.44-mg/cm$^2$-thick aluminum-foil shield was placed between the target and the telescope to
prevent fission fragments and $\delta$ electrons produced in the target from
damaging the $\Delta E$ detector and degrading detector performance.
Fission fragments were detected with a third 146-$\mu$m-thick Micron S2 silicon detector located 19 mm upstream
from the target.
The silicon detectors are segmented into 48 0.5-mm-wide rings on one side and 16 22.5\degree-wide sectors
on the other.  For this experiment, pairs of adjacent rings
and sectors were bussed together to form 24 1-mm-wide rings and 8 45\degree-wide sectors.
The silicon detectors are also coated with 27-$\mu$g/cm$^2$ aluminum contacts on the ring side and
500-$\mu$g/cm$^2$ gold contacts on the sector side.
The gold can significantly straggle the fission fragments, making energy separation between scattered $\alpha$ particles
and fission fragments difficult. To minimize straggling, the fission detector was installed with the ring side facing
downstream and the $^{240}$Pu target was mounted with the electroplated surface facing upstream. 

The target wheel and silicon detectors were mounted inside a vacuum chamber, which was surrounded by
a neutron detector (referred to as ``NeutronBall'') consisting of a tank filled with 3.5 tons of liquid scintillator.
The tank is segmented into six regions: four identical quadrants that make up the central cylinder and two endcaps.
Twenty photomultiplier tubes (PMTs), three on each quadrant and four on each endcap, are used to measure scintillation light.
At the time of the measurement, the central cylinder was filled with fresh EJ-335 liquid scintillator doped with 0.25-wt\% of 
natural gadolinium \cite{EljenTechnology}; however the two endcaps contained degraded liquid scintillator with poor 
optical transmission.  Therefore, in the present work, only events detected by the twelve PMTs on the central cylinder 
were included in the data analysis.  

\subsection{Detector calibrations}
\label{subsec:DetectorCalibrations}
For the $\Delta E$ and $E$ detectors, the response of each ring and sector was calibrated with a $^{226}$Ra $\alpha$
point source that provided the following $\alpha$ lines: 4784, 5304, 5489, 6002, and 7687 keV \cite{NNDC}.
At 7687 keV, the resulting 1$\sigma$ energy resolutions for the $\Delta E$ detector and $E$ detector were
approximately 40 keV and 24 keV, respectively. 
The fission detector was calibrated with a $^{252}$Cf spontaneous fission source.
The light and heavy fission-product mass peaks were used to gain match the response of the rings.  
For NeutronBall, $^{60}$Co and $^{228}$Th $\gamma$-ray point sources provided
calibration points at 1253 keV (the average energy of the 1173-keV and 1332-keV $\gamma$~rays from $^{60}$Co)
and 2615 keV (from $^{208}$Tl in the $^{228}$Th decay chain) \cite{NNDC}.
Another calibration point was provided by the 4440-keV $\gamma$ rays \cite{NNDC} that were emitted
following inelastic $\alpha$ scattering with the natural-carbon target that promoted $^{12}$C
to its first excited state. The energy resolution of the liquid scintillator at energy $E$ (in MeV)
was $\sigma(E)/E = 25\%/\sqrt{E}$~\cite{Akindele2017}.

The efficiency for detecting a single neutron with the central cylinder of NeutronBall was determined to be 0.504(5)
and was measured by placing a $^{252}$Cf fission source at the target position. More details will be given in
Sec.~\ref{subsec:AveragePromptFissionNeutronMultiplicity}.

\subsection{$\alpha$-particle beam}
\label{subsec:AlphaParticleBeam}
The $\alpha$-particle beam-spot size was approximately 3~mm in diameter and was observed with an in-vacuum camera
that imaged the phosphor targets.
The exact beam energy provided by the K150 Cyclotron was determined from data collected for the $^{208}$Pb target.
Scattering of $\alpha$ particles to discrete states in $^{208}$Pb was used as an \textit{in situ} calibration.
The beam energy was determined to be 53.9(1) MeV. This value allowed 
the excitation energy of the $^{208}$Pb nucleus to be properly reconstructed after taking into account
the energy deposition in the $\Delta E$-$E$ telescope, the energy loss in dead layers
(i.e., the target, the aluminum-foil shield, and the gold and aluminum contacts on the
surfaces of the silicon detectors), and the recoil energy of the $^{208}$Pb nucleus. 
The uncertainty in the beam energy was taken to be the 1$\sigma$ width of the $\alpha$ peak corresponding
to elastic scattering.

\section{Analysis and results}
\label{sec:AnalysisAndResults}
A $^{240}$Pu($\alpha$,$\alpha^{\prime}f$) interaction was indicated by a coincidence between an
$\alpha$ particle hitting the silicon telescope and a fission fragment hitting the fission detector.
For a $^{240}$Pu CN with excitation energy $E_\text{ex}$, corresponding to an equivalent incident
neutron energy $E_n$, the average prompt-fission-neutron multiplicity was determined from

\begin{equation}
    \bar{\nu}(E_n) = \frac{N_{n}(E_n)}{N_{\alpha-f}(E_n)\epsilon_n}, 
\label{eq:Nubar}
\end{equation}
where $N_{\alpha-f}(E_n)$ is the number of measured $^{240}$Pu($\alpha$,$\alpha^{\prime}f$) $\alpha$-fission coincidences
at $E_n$, $N_n(E_n)$ is the number of detected prompt fission neutrons associated with these coincidences, and
$\epsilon_n$ is the single-neutron detection efficiency for the central cylinder of NeutronBall.
The analysis performed to obtain the quantities in Eq.~\ref{eq:Nubar} is discussed in this section,
and the resulting $\bar{\nu}(E_n)$ distribution is given.

\subsection{Particle identification and event selection}
\label{subsec:ParticleIdentificationAndEventSelection}

\subsubsection{Charged particles}
\label{subsubsec:ChargedParticles}
\begin{figure}[!tb]
   \includegraphics[width=0.5\textwidth]{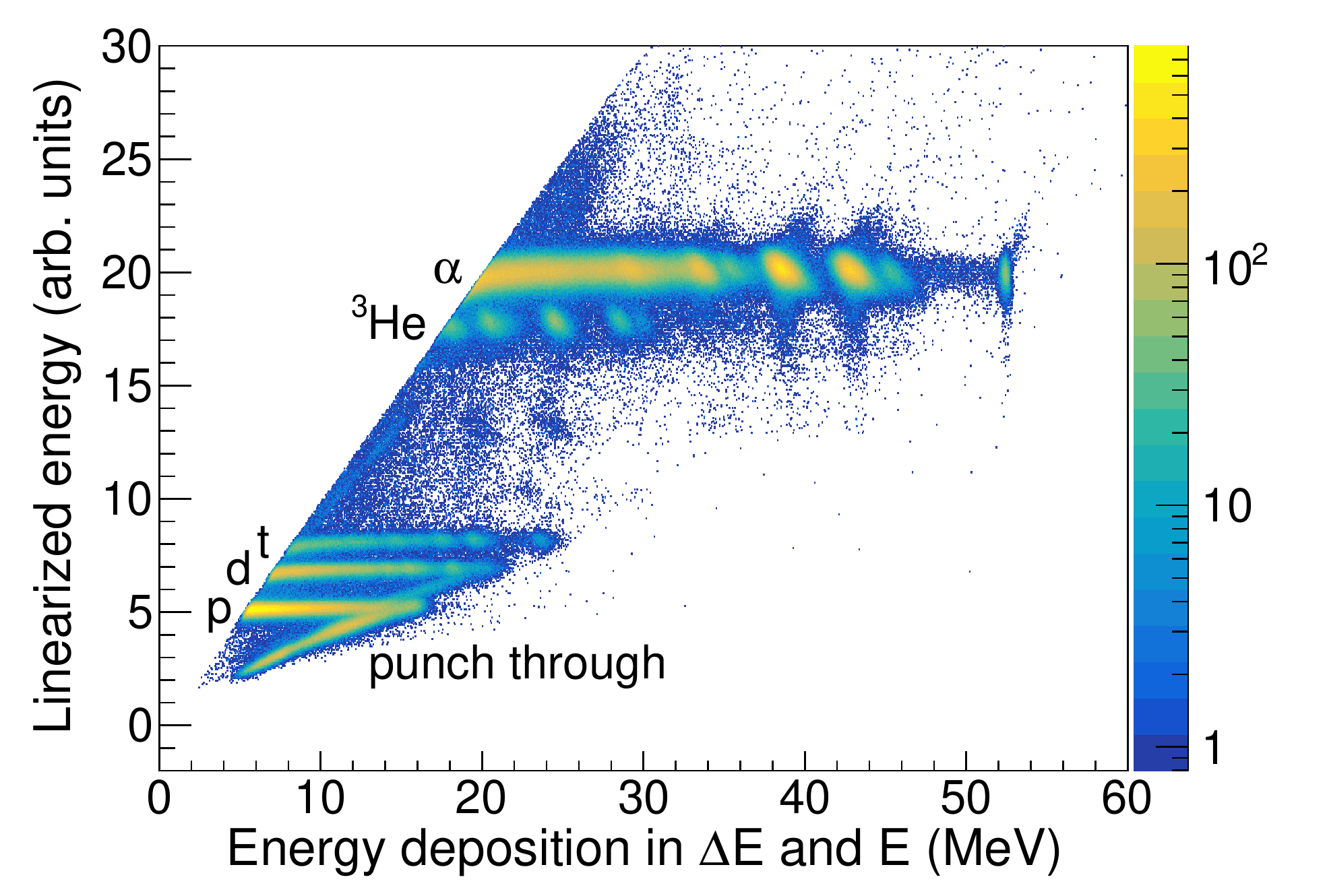}
   \caption
      {
      \label{fig:LinearizedPID}
      (Color online) Particle-identification plot for 53.9-MeV $\alpha$ particles incident on $^{240}$Pu.
      The linearized energy versus the total energy deposited in both the $\Delta E$ and $E$ detectors is shown
      for events hitting a chosen ring in the $\Delta E$ detector.
      Bands corresponding to protons (p), deuterons (d), tritons (t), $^3$He, and $\alpha$ particles ($\alpha$)
      are indicated. The diagonal streaks are due to high-energy charged particles that ``punch through'' the
      $E$ detector and therefore do not deposit all of their energy in the telescope.
      }
\end{figure}

For events in the silicon telescope, the energies deposited in the $\Delta E$ and $E$ detectors
($E_{\Delta E}$ and $E_{E}$, respectively) were used for particle identification (PID).
Protons, deuterons, tritons, $^{3}$He, and $\alpha$ particles were distinguished by plotting
the ``linearized energy'' $E_\text{lin}$~\cite{Goulding1964} versus 
the total energy deposition in both the $\Delta E$ and $E$ detectors, where

\begin{equation}
   E_\text{lin} = [(E_{\Delta E} + E_{E})^{1.75} - E_{E}^{1.75}]^{1/1.75}. 
\label{eq:Linearization}
\end{equation}
Alpha-particle events were isolated by generating a PID plot for each $\Delta E$-detector ring
(e.g., Fig.~\ref{fig:LinearizedPID}) and gating on the region above $^3$He ($E_\text{lin}$
approximately between 16.5--24).

\subsubsection{Fission}
\label{subsubsec:Fission}
\begin{figure}[!tb]
   \includegraphics[width=0.5\textwidth]{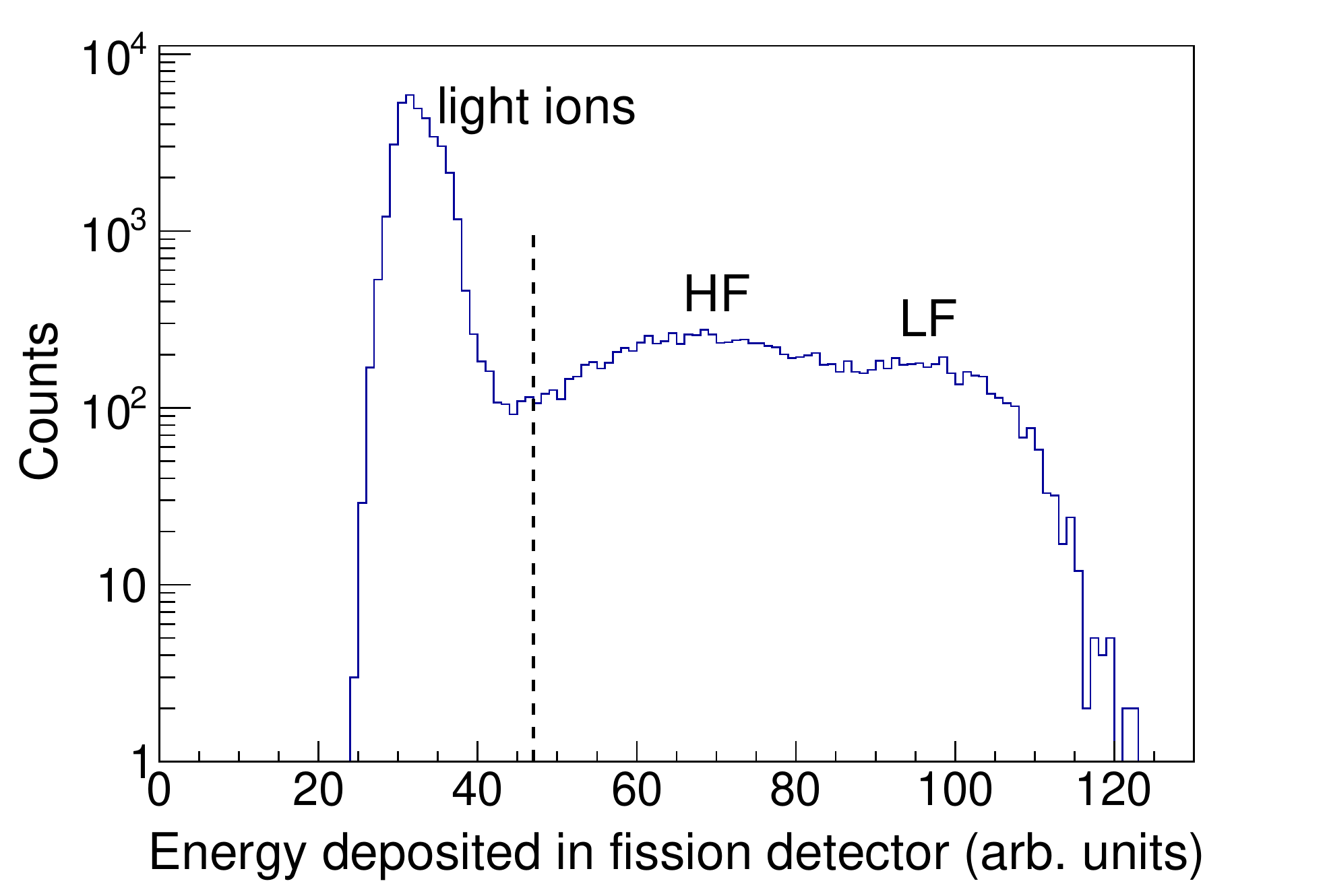}
   \caption
      {
      \label{fig:FissionSpectrum}
      Gain-matched spectrum measured by a single ring on the fission detector for 53.9-MeV $\alpha$
      particles incident on $^{240}$Pu. Peaks corresponding to heavy fission fragments (HF),
      light fission fragments (LF), and light ions are labeled. A vertical line is drawn at the
      energy cut used to separate fission fragments and light ions.  
      }
\end{figure}

Fig.~\ref{fig:FissionSpectrum} shows the gain-matched spectrum measured by a single ring on the fission detector
for $\alpha$ particles incident on the $^{240}$Pu target. A double hump is present at higher energies due to heavy and
light fission fragments hitting the detector.
The large peak at lower energies is primarily due to light ions from $^{240}$Pu $\alpha$ decay and
$\alpha$-particle interactions with carbon and oxygen in the $^{240}$Pu target, which was confirmed by analysis of the
data collected for the natural-carbon and Mylar targets.  
For each ring, fission events were selected and light-ion events removed by cutting above an energy deposition of
47 (arb. units).

\subsubsection{$^{240}$Pu($\alpha$,$\alpha^{\prime}f$) events}
\label{subsubsec:240PuInelasticAlphaScatteringEvents}
\begin{figure}[!tb]
   \includegraphics[width=0.5\textwidth]{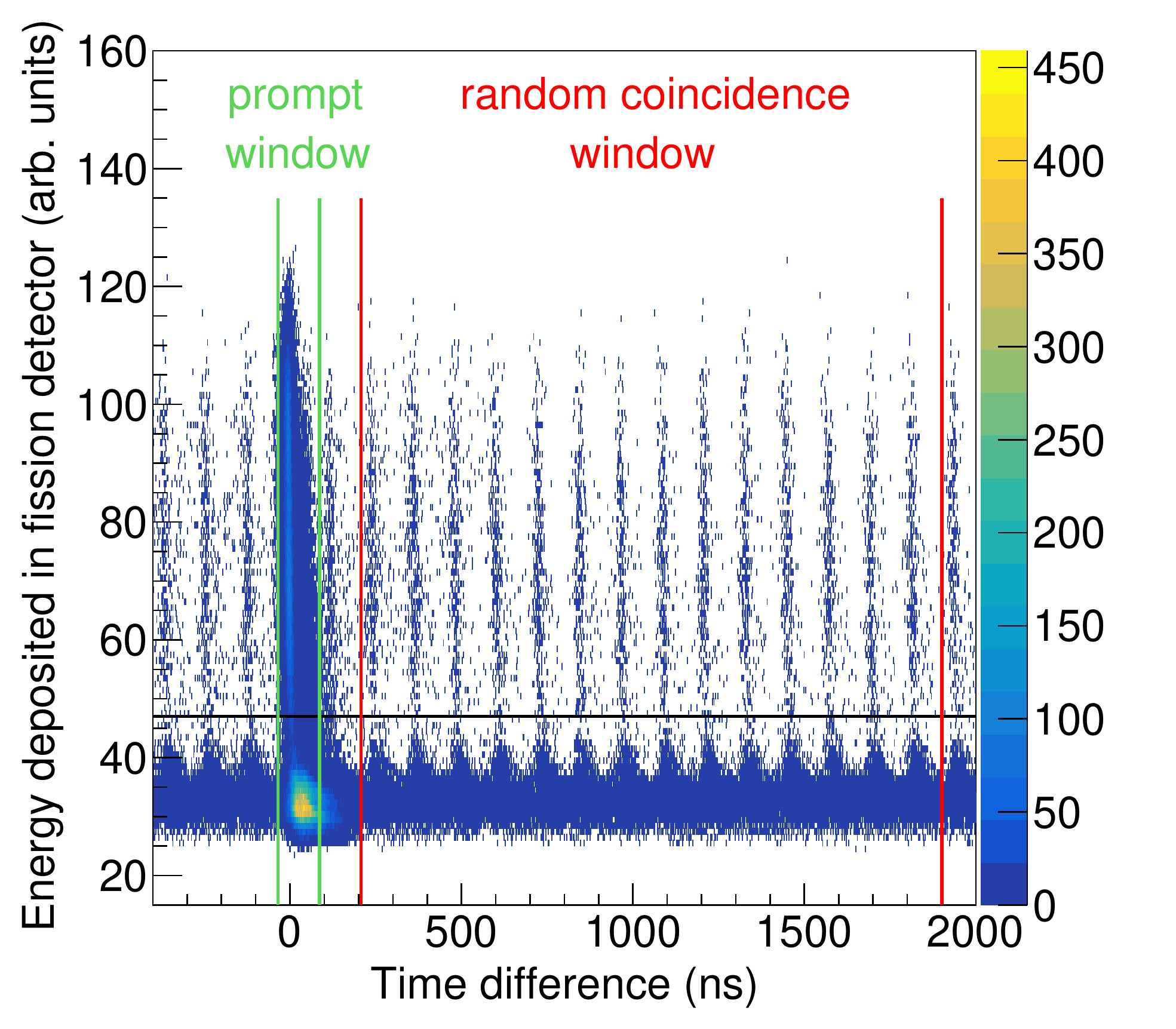}
   \caption
      {
      \label{fig:FissionSiliconCoincidencePlot}
      (Color online) Coincidences between an $\alpha$ particle hitting the $\Delta E$-$E$ telescope and an event
      in the fission detector. The energy deposited in the fission detector is plotted versus the fission-detector
      event time minus the $\alpha$-particle event time. A horizontal line is drawn at the energy cut-off used
      to isolate fission fragments from light ions. Vertical lines indicate the gates used in the data analysis
      to identify and characterize prompt $\alpha$-fission coincidences ($-35$~ns to 86~ns) and random-coincidences
      (207 to 1901~ns). 
      }
\end{figure}

In Fig.~\ref{fig:FissionSiliconCoincidencePlot}, the time difference between coincident $\Delta E$-$E$
$\alpha$-particle and fission-detector events is plotted; the energy deposited in the fission detector
is given along the \textit{y} axis.  A horizontal line is drawn at the energy cut-off used to isolate
fission fragments from light ions.
Coincidences above the cut-off with a time difference between $-35$~ns and 86~ns (``prompt'' region)
were tagged as $^{240}$Pu($\alpha$,$\alpha^{\prime}f$) events.
The small bursts of events present every 121~ns in Fig.~\ref{fig:FissionSiliconCoincidencePlot} coincide with
the K150 cyclotron frequency and are due to random coincidences such as an $\alpha$ particle hitting the
$\Delta E$-$E$ telescope and a fission fragment from a $^{240}$Pu($\alpha$,$f$) reaction in the target
hitting the fission detector.

\subsubsection{Neutrons}
\label{subsubsec:Neutrons}
PMT signals that arrived within a coincidence window of 200~ns were assumed to come from
a single event in NeutronBall, e.g., a neutron capture on gadolinium, or an interaction of a
room-background $\gamma$~ray.  These signals were first gain matched, as described in Ref.~\cite{Akindele2017},
then summed together to acquire the total energy deposited by the event.
Only events with energy greater than 2 MeV were included in the data analysis to exclude 
most of the contribution from backgrounds and electronic noise.

\begin{figure}[!tb]
   \includegraphics[width=0.5\textwidth]{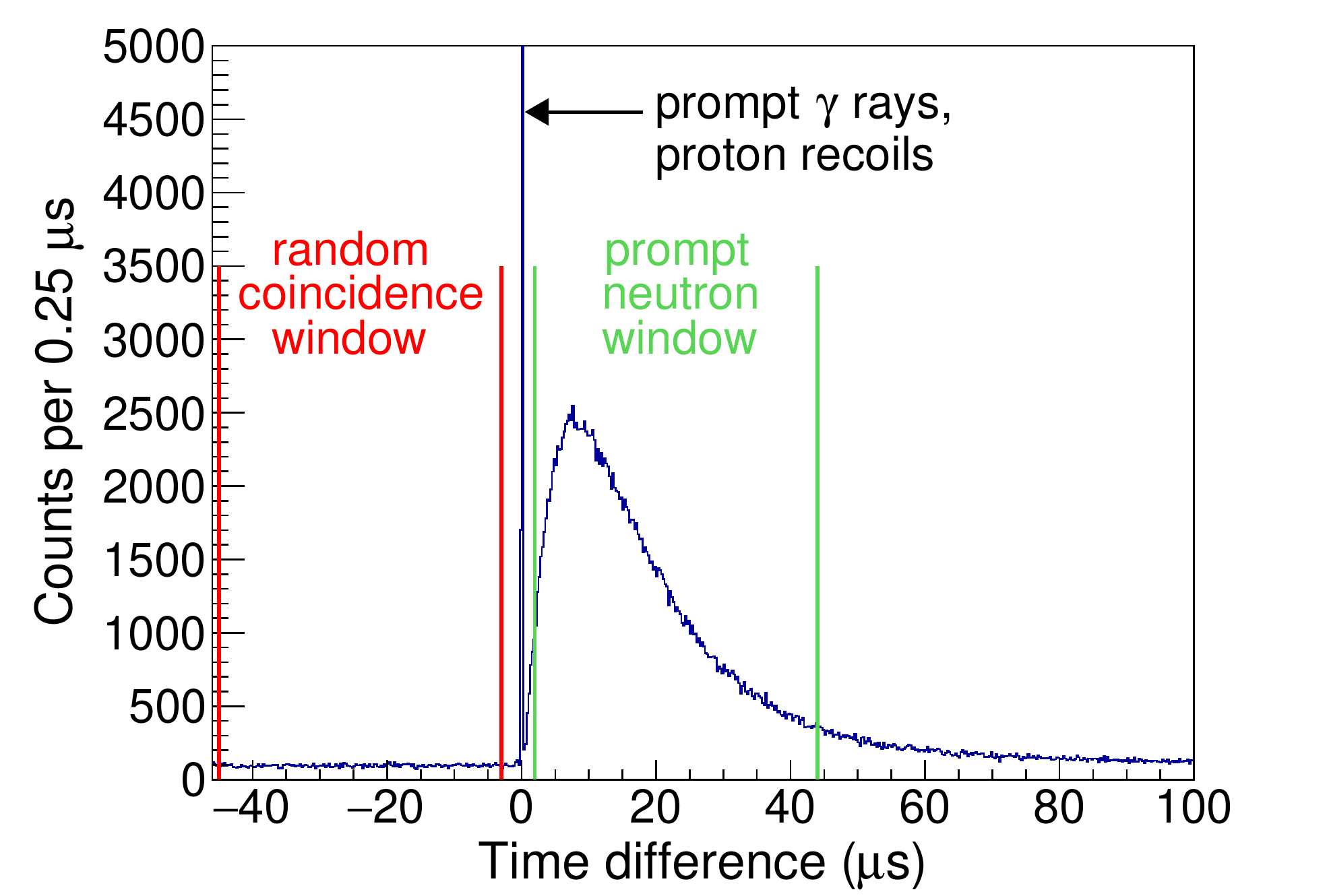}
   \caption
      {
      \label{fig:TimeDifferenceSpectrum_Neutrons}
      (Color online) Time difference between an event in NeutronBall and an $\alpha$-fission coincidence tagged as a
      $^{240}$Pu($\alpha$,$\alpha^{\prime}f$) event. The peak due to prompt fission $\gamma$ rays and
      proton recoils is indicated. The time windows used in the analysis to gate on prompt fission
      neutrons (2~to~44~$\mu$s) and random coincidences ($-45$~to~$-3$~$\mu$s) are also shown. 
      }
\end{figure}

For the tagged $^{240}$Pu($\alpha$,$\alpha^{\prime}f$) events, a timing gate was opened
50~$\mu$s before and closed 500~$\mu$s after the $\alpha$-fission coincidence. 
The time difference between a NeutronBall event occurring within this gate and the
$\alpha$-fission coincidence was plotted (Fig.~\ref{fig:TimeDifferenceSpectrum_Neutrons}).
The sharp peak around 0~$\mu$s in Fig.~\ref{fig:TimeDifferenceSpectrum_Neutrons} is from the
flash of prompt $\gamma$~rays following fission and from proton recoils generated during thermalization
of the neutron in the liquid scintillator.
The broad peak above 0~$\mu$s is attributed to prompt fission neutrons;
its width is determined by the moderation time of the neutrons in the scintillator.
Both features lie on top of a flat background due to random coincidences. 

\subsection{Equivalent neutron energy}
\label{subsec:EquivalentNeutronEnergy}
\begin{figure}[!tb]
   \includegraphics[width=0.5\textwidth]{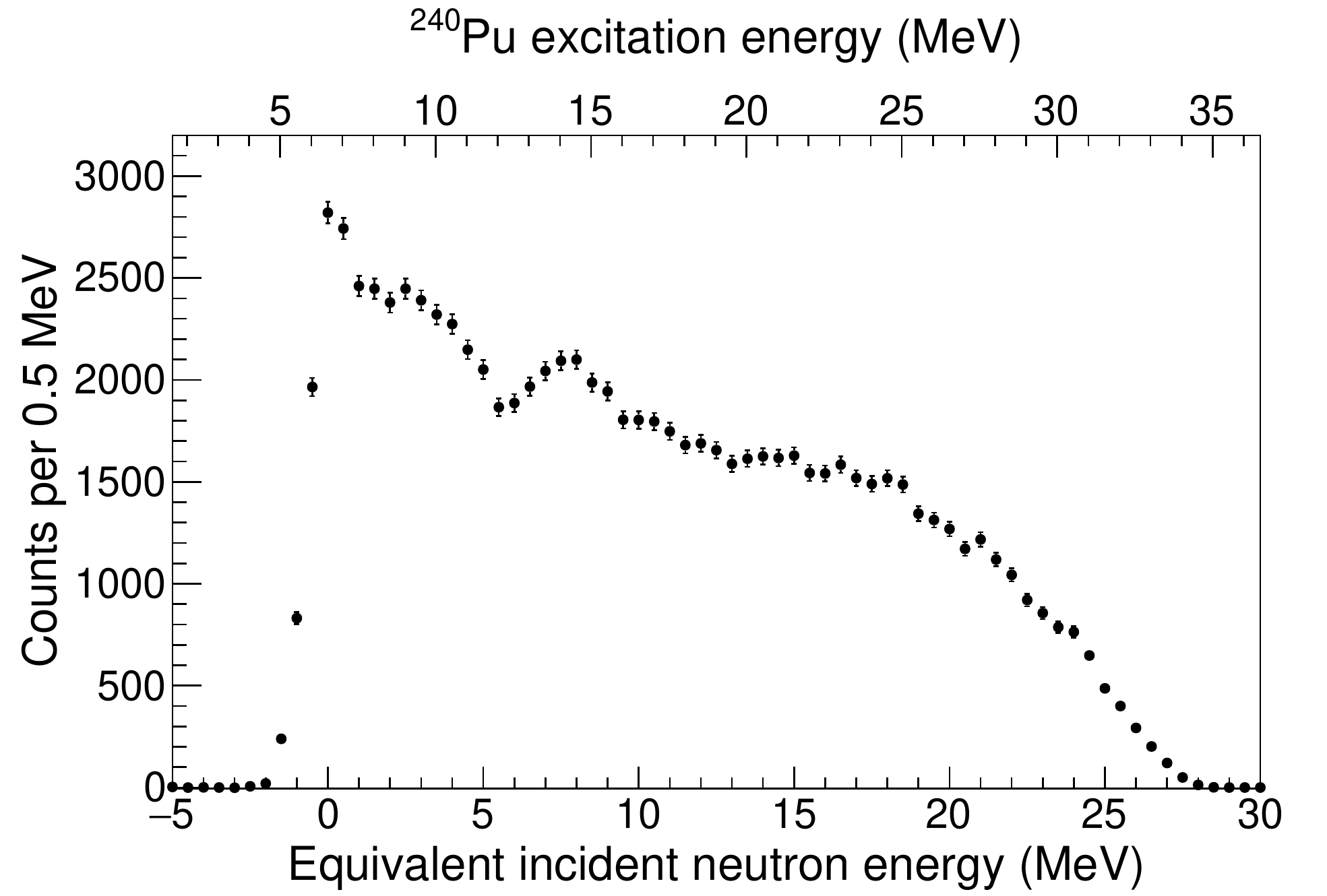}
   \caption
      {
      \label{fig:EquivalentNeutronEnergy}
      The distribution of equivalent incident neutron energies (and corresponding $^{240}$Pu excitation energies)
      for $^{240}$Pu($\alpha$,$\alpha^{\prime}f$) events; 0.5-MeV-wide energy bins are used.
      }
\end{figure}

The excitation energy $E_\text{ex}$ of $^{240}$Pu following inelastic $\alpha$-particle scattering
was determined from the beam energy $E_{\alpha}$, the scattered-$\alpha$-particle energy
$E_{\alpha^{\prime}}$, and the $^{240}$Pu recoil energy $E_r$:

\begin{equation}
    E_\text{ex} = E_{\alpha} - E_{\alpha^{\prime}} - E_r. 
\label{eq:NuclearExcitationEnergy}
\end{equation}
The value of $E_{\alpha^{\prime}}$ was the total energy deposited in the $\Delta E$-$E$ telescope
corrected for energy losses in the target, the $\delta$ shield, and the inert gold and aluminum contacts on the
surfaces of the silicon detectors.  The equivalent incident neutron energy $E_n$ was then determined from

\begin{equation}
    E_n = \frac{m_t+m_n}{m_t}(E_\text{ex} - S_n), 
\label{eq:EffectiveIncidentNeutronEnergy}
\end{equation}
where $m_t$ is the mass of $^{239}$Pu, $m_n$ is the neutron mass, and $S_n$ is the neutron separation energy
for $^{240}$Pu. 
Fig.~\ref{fig:EquivalentNeutronEnergy} shows the $E_n$ distribution for $^{240}$Pu($\alpha$,$\alpha^{\prime}f$) events.
The corresponding $^{240}$Pu excitation energy is also given.
Fission of $^{240}$Pu starts to occur at ${E_n=-1.61}$~MeV (4.9-MeV $^{240}$Pu excitation energy) \cite{Northrop1959}.
The feature at ${E_n\sim5.5}$~MeV is due to $^{240}$Pu second-chance fission \cite{Lestone2014, Stetcu2016},
and above ${E_n\sim18.5}$~MeV, the number of events tapers off quickly due to the $\alpha$-Pu Coulomb barrier. 

\subsection{Average prompt-fission-neutron multiplicity}
\label{subsec:AveragePromptFissionNeutronMultiplicity}
The average prompt-fission-neutron multiplicity was obtained with Eq.~\ref{eq:Nubar} for equivalent incident
neutron energies ranging between 0.25 and 26.25 MeV.  
The quantity $N_{\alpha-f}(E_n)$ in Eq.~\ref{eq:Nubar} is the number of $\alpha$-fission coincidences in the
(121-ns-wide) prompt region of Fig.~\ref{fig:FissionSiliconCoincidencePlot}, corrected for the contribution from
random coincidences.  This contribution was determined by taking the sum of $\alpha$-fission coincidences
in the region 207--1901~ns and scaling down to a 121-ns-wide time window.

The number of neutrons $N_n(E_n)$ was obtained by taking the difference between the total counts in the
time regions 2~to~44~$\mu$s and $-45$~to~$-3$~$\mu$s in Fig.~\ref{fig:TimeDifferenceSpectrum_Neutrons}.
The contribution from random $\alpha$-fission coincidences was determined from
the time-difference spectrum for NeutronBall events associated with the 207--1901-ns region in
Fig.~\ref{fig:FissionSiliconCoincidencePlot} (scaled down to correspond to a 121-ns-wide
$\alpha$-fission time window). 

A single-neutron detection efficiency of ${\epsilon_{n} = 0.504(5)}$ was obtained by first recording
the time-difference between $^{252}$Cf fission events in the fission detector and events in NeutronBall. 
The total number of prompt neutrons measured was then determined and divided by the number
of fission events and $\bar{\nu}$ for $^{252}$Cf (i.e., 3.757) \cite{Boldeman1985, Boldeman1967}. 

\begin{figure}[!tb]
   \includegraphics[width=0.5\textwidth]{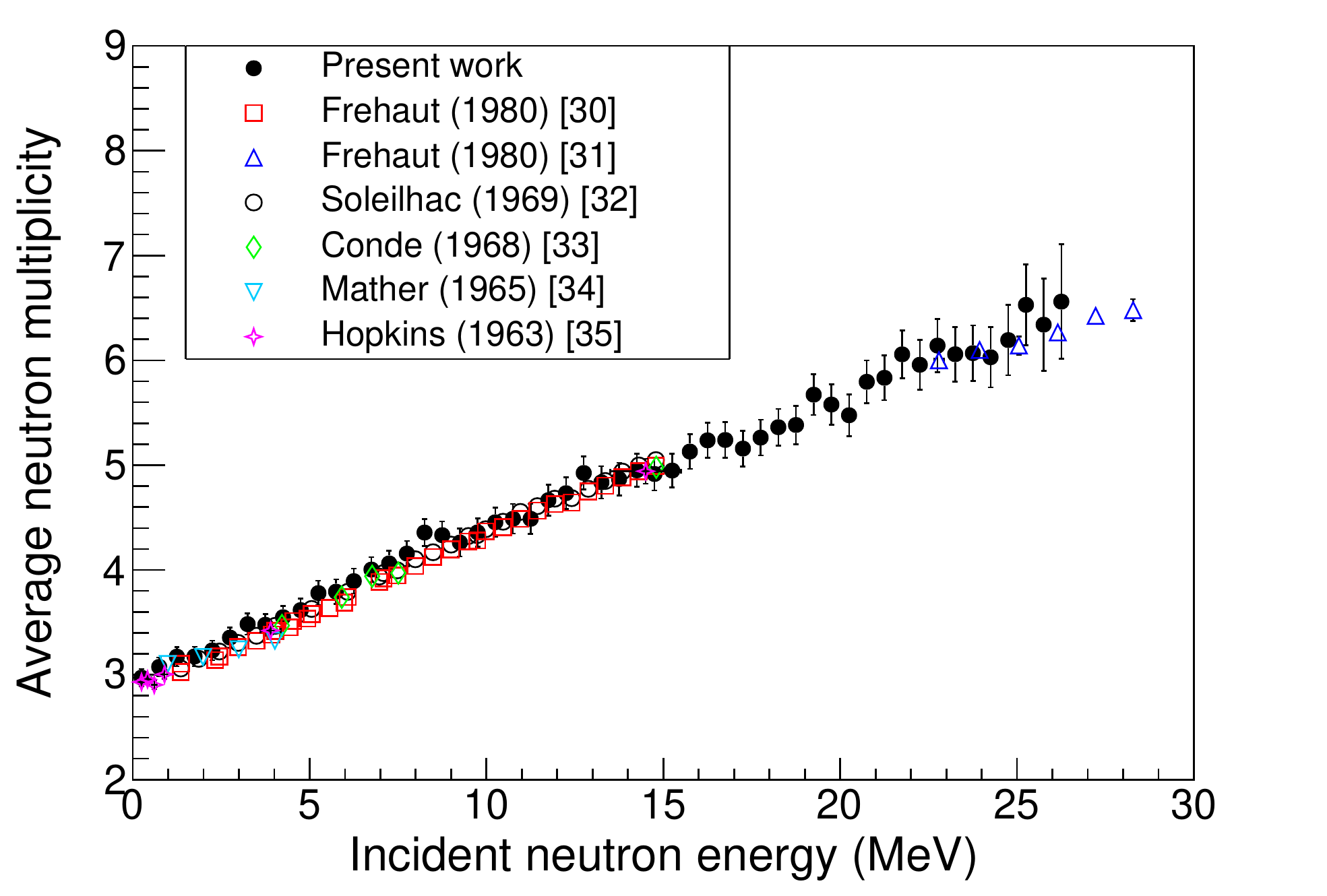}
   \caption
      {
      \label{fig:NubarVersusNeutronEnergy}
      (Color online) The average prompt-fission-neutron multiplicity $\bar{\nu}$ as a function of incident neutron energy
      for $^{239}$Pu($n$,$f$).     
      In the present work, $\bar{\nu}$ has been determined continuously from 0.25--26.25~MeV in 0.5-MeV-wide intervals.
      The results are compared with direct neutron measurements found in literature~\cite{Frehaut1980_1, Frehaut1980_2, Soleilhac1969, Conde1968, Mather1965, Hopkins1963}.
      In the present work the uncertainties are primarily due to counting statistics; for the literature values,
      most of the uncertainties are smaller than the data markers.
      }
\end{figure}

The $\bar{\nu}(E_n)$ distribution obtained is given in Fig.~\ref{fig:NubarVersusNeutronEnergy}.
Each $\bar{\nu}$ value and its uncertainty is also provided in Table~\ref{tab:Nubar};
the uncertainty is dominated by the statistical uncertainties in the number of $\alpha$-fission
coincidences and the number of detected neutrons.
Fig.~\ref{fig:NubarVersusNeutronEnergy} also shows that the results of the present work are consistent with 
direct neutron measurements for $^{239}$Pu($n$,$f$)~\cite{Frehaut1980_1, Frehaut1980_2, Soleilhac1969, Conde1968, Mather1965, Hopkins1963},
providing validation that the surrogate-reaction method can be used to determine $\bar{\nu}$ for actinides.

\begin{table}[!b]
\caption{\label{tab:Nubar} Equivalent incident neutron energies $E_n$ and corresponding
        $\bar{\nu}$ values from the present work.
	} 
\begin{ruledtabular}
\begin{tabular}{ d c | d c }
  \multicolumn{1}{c}{$E_n$ (MeV)} & \multicolumn{1}{c|}{$\bar{\nu}$} & \multicolumn{1}{c}{$E_n$ (MeV)} & \multicolumn{1}{c}{$\bar{\nu}$}\\ 
  \hline\noalign{\smallskip}
  0.25 & 2.97(8) & 13.75 & 4.87(15) \\
  0.75 & 3.08(9) & 14.25 & 4.95(16) \\
  1.25 & 3.17(9) & 14.75 & 4.92(16) \\
  1.75 & 3.17(9) & 15.25 & 4.95(16) \\
  2.25 & 3.23(9) & 15.75 & 5.13(17) \\
  2.75 & 3.35(10) & 16.25 & 5.24(17) \\
  3.25 & 3.48(10) & 16.75 & 5.24(17) \\
  3.75 & 3.48(10) & 17.25 & 5.16(17) \\
  4.25 & 3.55(11) & 17.75 & 5.26(17) \\
  4.75 & 3.62(11) & 18.25 & 5.36(17) \\
  5.25 & 3.78(12) & 18.75 & 5.38(18) \\
  5.75 & 3.79(12) & 19.25 & 5.67(19) \\
  6.25 & 3.89(12) & 19.75 & 5.58(19) \\
  6.75 & 4.00(12) & 20.25 & 5.48(20) \\
  7.25 & 4.06(12) & 20.75 & 5.80(20) \\
  7.75 & 4.16(12) & 21.25 & 5.83(21) \\
  8.25 & 4.36(13) & 21.75 & 6.06(23) \\
  8.75 & 4.33(13) & 22.25 & 5.96(24) \\
  9.25 & 4.26(13) & 22.75 & 6.14(25) \\
  9.75 & 4.36(13) & 23.25 & 6.06(26) \\
 10.25 & 4.46(14) & 23.75 & 6.07(27) \\
 10.75 & 4.49(14) & 24.25 & 6.03(29) \\
 11.25 & 4.49(14) & 24.75 & 6.19(34) \\
 11.75 & 4.67(15) & 25.25 & 6.53(39) \\
 12.25 & 4.73(15) & 25.75 & 6.34(44) \\
 12.75 & 4.93(16) & 26.25 & 6.56(55) \\
 13.25 & 4.84(15) & & \\
\end{tabular} 
\end{ruledtabular}
\end{table}

\section{Summary and Conclusions}
\label{sec:SummaryAndConclusions}

$^{240}$Pu($\alpha$,$\alpha^{\prime}f$) was used as a surrogate reaction to determine the $^{239}$Pu($n$,$f$)
prompt-fission-neutron multiplicity as a function of incident neutron energy from 0.25--26.25~MeV.
This is the first time $\bar{\nu}$ for $^{239}$Pu($n$,$f$) has been obtained continuously over this neutron
energy range in a single measurement.
The results of the present work are in good agreement with those from direct neutron
measurements~\cite{Frehaut1980_1, Frehaut1980_2, Soleilhac1969, Conde1968, Mather1965, Hopkins1963}.
Similar conclusions were drawn in Ref.~\cite{Akindele2019}, where the surrogate reaction
$^{242}$Pu($\alpha$,$\alpha^{\prime}f$) was used to determine $\bar{\nu}$ for $^{241}$Pu($n$,$f$).
The success of these two experiments opens the door to using surrogate reactions to obtain $\bar{\nu}(E_n)$ for a
whole host of short-lived actinides that are currently inaccessible via direct methods.

\section{Acknowledgments}
We thank the staff of the Texas A\&M Cyclotron Institute for facilitating operations and facilities needed to
perform this measurement. This work was performed under the auspices of
the U.S. Department of Energy National Nuclear Security Administration
by Lawrence Livermore National Laboratory under Contract No. DE-AC52-07NA27344,
under Award No. DE-NA0000979,
and through the Nuclear Science and Security Consortium under Award No. DE-NA-0003180.

\bibliography{Bibliography}

\end{document}